\def\rem#1{}

\def\be{\begin{equation}}
\def\ee{\end{equation}}
\def\bea{\begin{eqnarray}}
\def\eea{\end{eqnarray}}

\documentclass[12pt]{article}
\usepackage[dvips]{graphicx}
\usepackage{amsmath,amssymb,bm}
\begin{document}
\title{\textbf{Superradiant instability for black holes immersed in a magnetic field}}
\author{R.A.Konoplya\\
\emph{Department of Physics, Kyoto University, Kyoto 606-8501, Japan}}

\date{}

\maketitle

\thispagestyle{empty}

\begin{abstract}
We consider the bound states of the massive scalar field around a rotating black hole immersed in the asymptotically uniform magnetic field.
In the regime of slow black hole rotation, the Klein-Gordon equation allows separation of variables.
We show that the growth rate of the instability can be amplified a few times by the magnetic field.
The effect occurs because the magnetic field adds the "effective mass" term $B |m|$ to the scalar field potential for a Kerr black hole.
In addition, and as a by-product, we discuss the behavior of the quasinormal modes for the magnetized rotating black holes.
\end{abstract}

\section{Introduction}

Interaction of black holes and magnetic fields is an important factor for large astrophysical black holes, and,
especially, for supermassive galactic black holes because of enormous magnetic fields in the nucleus of galaxies and near supermassive black holes.
Strong magnetic fields are induced also by accretion disks of rotating charged matter near black holes.
Yet, interaction of strong magnetic fields and black holes is, apparently, not limited just by the context of large scale astrophysical systems. Recent development in the  brane world theories opens possibility of observing quantum gravity at the Tev scale, so that mini black holes might be created at particle collisions in the Large Hadron Collider. Observing black holes in laboratories could make it possible to test interaction of black holes with external fields. After all, in the early universe strong magnetic fields could effect primordial black holes.

Sufficiently strong magnetic field near a black hole deforms the black hole space-time, so that one cannot consider the magnetic field as a test field on the black hole background, but a consistent solution of the Einstein-Maxwell equations is necessary instead. Fortunately, exact solutions for such a situation exist for both non-rotating (the Ernst solution) \cite{8} and rotating cases (the Diaz solution) \cite{9}.

Yet, the essential question is whether such a strong magnetic field, that could deform the black hole space-time, can exist, and, whether the Ernst and Diaz solutions can be considered as at least some approximate approach to a realistic situation?. The Ernst solution is described by the metric
\begin{equation}
d s^{2} = \Lambda^{2} \left(
\left(1- \frac{2 M}{r} \right) d t^{2} - \left(1- \frac{2 M}{r} \right)^{-1} d
r^{2} -r^{2} d \theta^{2}
\right)  - \frac{r^{2} \sin^{2} \theta}{\Lambda^{2} } d \phi^{2},
\end{equation}
where the external magnetic field is determined by the parameter $B$,
\begin{equation}
\Lambda = 1 + \frac{1}{4} B^{2} r^{2} \sin^{2} \theta,
\end{equation}
and the "unit" magnetic field measured in $Gs$ is
\begin{equation}
B_{M} = 1/M = 2.4 \times 10^{19} \frac{M_{Sun}}{M}.
\end{equation}

The magnetic field in the Ernst solution is poloidal and homogeneous far from the black hole.  This is not a big restriction for our consideration.
Indeed, although the large scale magnetic field in the observable universe has both toroidal and poloidal components, it is the poloidal component
that dominate in centers of galaxies and thereby in the region near a super-massive black hole. For mini black holes, which according to the brane-world scenarios, are expected to be observed in particles collisions, the homogeneous (poloidal) magnetic field is natural. Even though the super-strong accelerating magnetic field is screened in the region of particles collisions, we can assume, at least in principle, that if mini black holes are created in a laboratory, one could "immerse" mini-black holes in the magnetic field with required properties.

In order to make estimations of possible influence of the magnetic field on the supermassive black holes we need the two parameters at hand:
the magnetic field parameter $B$ and the mass of the black hole $M$. Modern observations suggest that super-massive black holes
in centers of galaxies can have mass $M \approx 10^6 - 10^9 M_{Sun}$, while observations of the magnetic field are much more complicated and lead sometimes to
controversial predictions for the value of the magnetic field in the centers of galaxies \cite{Zakharov:2002cf}, \cite{Contra}.
At the same time, observations of the Active Galactic Nuclei (AGN)
and micro- quasars indicate the existence of wide X-ray emission of lines of heavy ionized elements in their spectra \cite{Zakharov:2002cf}. This effect is
very well described when supposing existence of the super-strong magnetic field $B \approx 10^{10}- 10^{11}$ $Gs$ \cite{Zakharov:2002cf}. In addition some
other theories \cite{Tolubaev} imply existence of a strong magnetic field in the AGN. From these data for $M$ and $B$ at hand, by formula (2), one can easily see
that for the heaviest galactic black holes, the magnetic field can be of order of fractions  of a "unity" ($1/M$), thereby effecting the black hole metric itself.

The Diaz and Ernst solutions, which we shall analyze here, describe the black holes immersed in a magnetic field decaying near the black hole to some asymptotic value, so that far from black hole the magnetic field is uniform and represents the Melvin Universe. Different effects around such black holes have been studied in \cite{10} and some generalizations were obtained in \cite{11}. Potentially observable effects, such as quasinormal modes and gravitational lensing were considered in \cite{12}, \cite{13},  \cite{14} for non-rotating magnetized black holes.
In this paper, we shall consider scattering of scalar waves around rotating magnetized black holes.
An incident wave near the black hole will be partially absorbed (tunneled through the potential barrier) and
partially radiated away as a response of a black hole to an external perturbation. When a black hole of mass $M$ and radius $r_{+}$
is rotating, and the real oscillation frequency of the perturbation (for a mode with an azimuthal number $m$) satisfies the inequality
\begin{equation}
\omega < \frac{a m}{2 M r_{+}},
\end{equation}
then the energy radiated away exceeds the energy of the incident wave. This is the well known effect of super-radiance predicted by Zeldovich and Misner \cite{15} and
computed for Kerr black holes by Starobinsky \cite{16}. Press and Teukolsky suggested that if the wave radiated away will be reflected by a mirror surrounding a black hole,
one could make a kind of "black hole bomb", because initial small perturbations would grow without bound \cite{17}.
The role of the mirror can play the potential well, another local minimum, which appears because of the non-vanishing massive term \cite{18}.

In the limit $M \mu <<1$, where $M$ is the black hole mass and
$\mu$ is the inverse Compton wavelength of the particle, Detweiler showed that the massive scalar field exerts the superradiant instability with the  maximal growth rate \cite{20}, \cite{21}
(at $a=M$) given by the formula
\begin{equation}
\gamma = \frac{1}{24} \frac{a}{M} (\mu M)^{9} (G M/c^{3})^{-1}.
\end{equation}
This instability has very small growth rate and probably cannot be significant for real black holes. Accurate calculations of instability
made recently in \cite{18_1} allowed to observe the maximal
growth
\begin{equation}
\gamma = 1.5 10^{-7} (G M/c^{3})^{-1}.
\end{equation}
at $M \mu =0.42$. Yet, due-to short lifetimes or large particle masses, this instability is tiny for any known massive boson particles, except
possibly, the neutral pion $\pi^{0}$ when $M \sim 10^{12} kg.$

In this paper we shall consider the bound states of massive scalar field in the vicinity of the magnetized black holes and analyze the corresponding superradiant instability.
As a by-product we shall consider the $B$-corrections to the orbital geodesic motion (as that for which the instability is important) of particles around the Ernst black hole. The paper is organized as follows. In Sec. II the massive scalar field equation is decoupled for the regime of "small" (in comparison with a "unity" $1/M$) magnetic field, and is reduced to the wave-like equation with an effective potential.
Sec. III considers the bound states and their instability in the limit $\mu M \ll 1$. In particular, we obtain the generalization of the well-known
Detweiler formula for the  superradiant instability for a non-zero external magnetic field.
In Sec. IV we briefly discuss the quasinormal modes of rotating magnetized black holes.
In the Conclusion we summarize the obtained results and outline the number of open questions.

\vspace{4mm}


\section{The metric and separation of variables in the wave equation.}

The Diaz metric is the exact solution of the Einstein-Maxwell equation and represents a charged rotating Kerr-Newman black holes immersed in an asymptotically
uniform electromagnetic field (se formulas (2-11) in \cite{19} and \cite{9}). If we put the black hole charge, and external electric field to be zero, then
the metric takes the form

\begin{equation}
d s^2 = |\Lambda|^2 \Sigma \left(\frac{\Delta}{A} d t^2 - \Delta^{-1} d r^2 - d \theta^2 \right) - \frac{A \sin^2 \theta}{\Sigma |\Lambda|^2 }(d \phi - W dt)^2.
\end{equation}

In order to separte variable in the Klein-Gordon equation, we have to be limited by the case of small magnetic fields $B << M^{-1}$, and small rotation $a << M$, so that we neglect terms of order $B^2 a^2$ or $B^4$ and higher. Within this approximation

\begin{equation}
\Lambda = 1 + \frac{1}{4} B^2 r^2 \sin^2 \theta + O(B^2 a^2, B^4),
\end{equation}

\begin{equation}
W = \frac{2 M r a}{A} + O(B^2 a^2, B^4),
\end{equation}

\begin{equation}
\Delta = r^2 + a^2 - 2 M r,
\end{equation}

\begin{equation}
\Sigma = r^2 + a^2 \cos^2 \theta,
\end{equation}

\begin{equation}
A = (r^2 + a^2)^2 - \Delta  a^2 \sin^2 \theta.
\end{equation}

Here $B$ is the parameter of the magnetic field, which gives the asymptotic value of the magnetic field. The solution (7), as was shown by Aliev and Galtsov \cite{Aliev1}, forms
the conical singularity  on a polar axis. This generates additional (to the Maxwell term) singular term in the stress-energy tensor on the right hand side of the Einstein equations. In order to avoid this singular term, one should consider the more general metric given in \cite{Aliev1} (see equation (2.9) in  \cite{Aliev1}). Yet in the slow rotation regime and zero background charge $Q=0$, the Aliev-Galtsov metric reduces to the uncharged Diaz metric, which we shall use here.

The massive scalar field obeys the equation

\begin{equation}
\frac{1}{\sqrt{-g}}\frac{\partial}{\partial x^{\mu}}\left(\sqrt{-g} g^{\mu \nu}  \frac{\partial \Phi}{\partial x^{\nu}}\right) = \mu^2 \Psi.
\end{equation}

As the background metric has the Killing vectors $(\partial_{\phi}, \partial_{t})$, it is implied that

\begin{equation}
\Phi \sim e^{i m \phi + i \omega t} S (\theta) R (r).
\end{equation}

The wave equation  (13) consists of the five terms, corresponding to components of (13) with $tt$, $rr$, $\theta \theta$, $t \phi$, and $\phi \phi$ derivatives.
The dependence on $\Lambda$ (and thereby on $B$) is preserved only in the last term $\phi \phi$, while in the other terms, $\Lambda^{-2}$
coming from the contra-variant components of the metric is canceled by  $\Lambda^{2}$ of $\sqrt{-g}$.
Neglecting small terms of orders $O(B^2 a^2, B^4, B^2 \mu^2)$ and higher, one can reduce the wave equation to
separate equations for $S (\theta)$ and $R (r)$ in the limit of small rotation and "weak" magnetic fields

\begin{equation}
\Delta \frac{d}{d r} \left( \Delta \frac{d R}{d r}  \right)+ (\omega^2 (r^2 + a^2)^2 - 4 a M r m \omega + a^2 m^2 - \Delta ((B^2 m^2 + \mu^2) r^2
+ a^2 \omega^2 + \lambda) )R = 0,
 \end{equation}

\begin{equation}
\frac{1}{\sin \theta}\frac{d}{d \theta}\left(\sin \theta \frac{d S(\theta)}{d \theta} \right) + (a^2 \omega^2 \cos^2 \theta - m^2 \sin^{-2}
\theta + \lambda) S(\theta).
\end{equation}

Thus we see that the radial equation has the form of wave equation for the Kerr black hole with the "effective mass" $\sqrt{\mu^2 + B^2 m^2}$,
while the equation for angular part looks like that for the massless scalar field, because the "massive" term in the angular part is of orders $a^2 B^2$.
Yet, that is not a problem for our future analysis, because it is enough to approximate $\lambda$ as $\ell (\ell +1)$. In the limit $a=0$ we have
the effective potential of the non-rotating case \cite{14}. It is worthwhile to note that in a similar fashion, the effective mass term appear for the massless scalar field when considering dominant term of the self-action of the scalar field or when a black hole is of Kaluza-Klein type \cite{Lemos}.

Let us stress here that the "small" magnetic field is a very good approximation to the real situation and means here in fact a super-strong magnetic field. Indeed, as the "unit" magnetic field measured in $Gs$ is $B_{M} = 1/M = 2.4 \times 10^{19} (M_{Sun}/M)$, and if we take $M =1$ for, say, a black hole of mass
$10^9 M_{Sun}$, then $B=1$ means $B = 2.4 \times 10^{10}$ $Gs$, what is about the possible maximum one can expect in astrophysical context \cite{Zakharov:2002cf}.
We made one implicit approximation, when discarding the terms of order $B^4$ and higher: in fact we discarded terms $B^4 r^4$ and higher, so that we can trust our equations (15, 16) only for $B r \ll 1$ ($M =1$). This again is not a serious restriction as even for a super-strong field, $B$ is a few orders less than unity
and thus the equations (15, 16) are a good approximation for $r$ ranging from the event horizon until $1/B$, what is few orders bigger than the radius of the event horizon.
In this way one has "infinity" situated somewhere around $r = 1/B$ and we solve thereby another problem: the exact Ernst or Diaz metrics are not asymptotically flat,
 because the magnetic field exists even at asymptotic infinity. This is certainly not physical situation and
the Ernst metric should have been matched somewhere far from the black hole to an asymptotically flat solution. We do not need to do it here because
the scattering processes happen near the peak of the effective potential barrier, and practically, incident wave comes to barrier not from asymptotic infinity but
from the far region $r \gg r_{g} = 2 M$. Therefore the scattering properties, such as absorption probabilities or quasinormal modes
are determined by the behavior of the effective potential near its maximum. In the astrophysical context, the independence of the quasinormal modes
on the asymptotic behavior of the effective potential was proved in \cite{KZ_PLB2}. (This certainly does not concern asymptotically AdS space-time where
the boundary conditions at spatial infinity is interpreted through the AdS/CFT correspondence).


\vspace{4mm}


\section{Bound states and superradiant instability}

The bound states appear when the effective potential has a local minimum. For $M=1$ and different values of $\ell$, bound states are formed for all values of the
effective mass term less than some threshold value $\sqrt{\mu^2 + B^2 m^2} < (\sqrt{\mu^2 + B^2 m^2})_{max}$. The values of  $(\sqrt{\mu^2 + B^2 m^2})_{max}$
for different values of $\ell$ are plotted in Fig. 1.
For large $\ell$ this maximal threshold value  is
\begin{equation}
(\sqrt{\mu^2 + B^2 m^2})_{max}  = \sqrt{\ell (\ell +1)}/2 \sqrt{3}, \quad (M=1)
\end{equation}

\begin{figure}[htbp]
\centering
\includegraphics[scale=0.5]{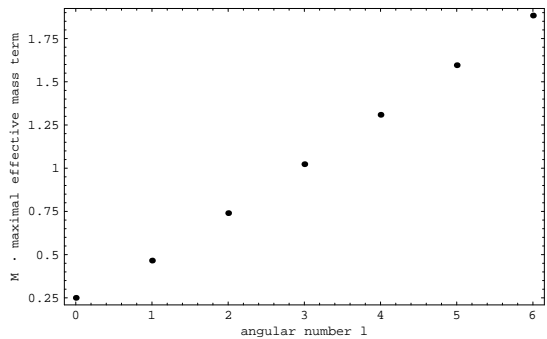}
\caption{The maximal value of the effective mass term $ M \sqrt{\mu^2 + B^2 m^2}$ allowing the bound metastable states for different values of $\ell$.
One can see that the maximal values of $M \sqrt{\mu^2 + B^2 m^2}$  approaches its asymptotic value $\sqrt{\ell (\ell +1)}/2 \sqrt{3} M$  in the regime of high
$\ell$ already at moderate $\ell$}
\label{1}
\end{figure}

From Fig. 1 one can see that this asymptotic regime is reached already at moderate
$\ell$. The discussion of the possible values of particle and black hole parameters when bound states are formed was given in \cite{thershold} for the Schwarzschild black hole.

The instability of the bound states is not very important if a particle moves towards a black hole: the particle will simply fall into the black hole. Yet for
particles moving around black holes on classically stable orbits, quantum metastability determines the time of orbiting around the black hole. Therefore
in the Appendix we shall find $B$ - corrections to the classical orbital motion of particles around Ernst black holes.

Let us start from the boundary conditions for the bound states problem
\begin{equation}\nonumber
\Psi \sim A e^{-i \omega r_{*}}, \quad r_{*} \rightarrow - \infty,
\end{equation}
\begin{equation}\nonumber
\Psi \sim 0, \quad r_{*} \rightarrow + \infty,
\end{equation}
which are the pure in-going waves at the event horizon and zero at the spatial infinity.
Under these boundary conditions the massive scalar field has bound states, which "live" in the local minimum of the effective potential barrier.

Here we shall follow the Detweiler approach which uses the approximation $\mu M <<1$. In our case,
in addition to this approximation, there are the two other restrictions: small rotation $a \ll 1$, and "small" magnetic field $B \ll 1$, (if $M=1$).
We shall not repeat all the steps of the Detweiler work \cite{20}, but just say that they are the same if one change $\mu^2$ by the $\mu^2 + B^2 m^2$.
This leads to the following result for the frequency of the bound state solution of the above wave equation,
\begin{equation}
\omega = (\sqrt{\mu^2 + B^2 m^2}) \sqrt{1- \left(\frac{(\sqrt{\mu^2 + B^2 m^2}) M}{\ell + 1 + n}\right)^{2}} + i \gamma,
\end{equation}
where the imaginary part $\gamma$ describes the rate of growth (or of decreasing) of each mode,

$$
\gamma = \sqrt{\mu^2 + B^2 m^2} (M(\sqrt{\mu^2 + B^2 m^2}))^{4 \ell + 4}(a m M^{-1} - 2 (\sqrt{\mu^2 + B^2 m^2}) r_{+}) \times
$$
$$
\frac{2^{4 \ell + 2}
(2 \ell +1 + n)!}{n! (\ell +1 + n)^{2 \ell + 4}} \left(\frac{\ell!}{(2 \ell)!
(2 \ell+1)!}\right)^{2} \times
$$
\begin{equation}
\Pi_{j=1}^{\ell} (j^2 (1-a^2/M^2) + ((am/M) - 2 (\sqrt{\mu^2 + B^2 m^2}) r_{+})^{2}).
\end{equation}

If one has $m>0$ and
\begin{equation}
a m M^{-1} - 2 (\sqrt{\mu^2 + B^2 m^2}) r_{+} > 0, \quad B \ll B_{M}
\end{equation}
then $\gamma >0$ and we have a super-radiant instability. The analog of the $2p$ state of the hydrogen atom ($\ell = m = 1$, $n=0$) has the fastest
instability growth, if we are limited by the regime $B M |m| << 1$. Due-to condition $\mu^2 B^2 \ll 1$, in the obtained formulas (18,19) we are limited by those values of the magnetic field $B$ and azimuthal number $m$, for which $B |m|$  at least a few times smaller than $\mu$, so that the contribution
due-to magnetic field plays a role of a relatively small (but not negligible) correction to the effective mass of the scalar field. In this case,
the "asymptotic" region $1/B$ is situated at a distance which is a few times larger than the distance from the black hole to the local minimum (see Fig. 2).

\begin{figure}[htbp]
\centering
\includegraphics[scale=0.6]{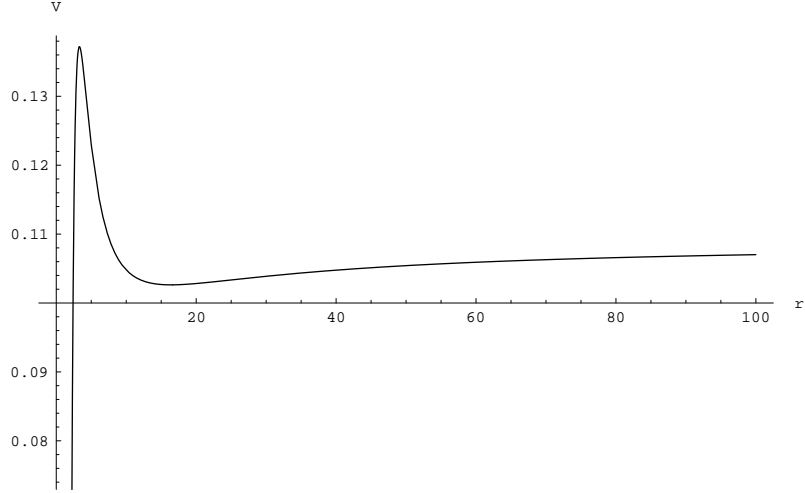}
\caption{The effective potential $m = \ell = 1$, $B =0.01$, $\mu = 0.3$, $M=1$.
One can trust the effective potential in the the region ($2$, $1/B$). The formal local minimum, where the bound states
can live, appears at $r \sim 20  M$.}
\label{1}
\end{figure}

The growth rate as a function of the magnetic field can be found in Fig. 3. There one can see that in the region of stability, the magnetic field increases the damping rate
of a mode, while in the region of superradiant instability, magnetic field can increase the growth rate by a few times. Let us stress that this result was obtained for
the slow rotation regime. It is well known that for highly rotating black holes the superradiant instability growth rate is increased by many orders \cite{19}, \cite{20}.
For highly rotating black holes the coupling between the projection of the angular momentum, given by $m$, and the magnetic field $B$, must be much stronger,
leading, probably, to a stronger superradiant instability. Anyway, let us speculate what happens if at least a similar, few times increasing
of the growth rate occurs for a highly rotating black holes.  For a superradiant instability to be significant in real processes it is important
that the e-folding time of the maximal instability growth be larger than each of the two quantities: the lifetime of the black hole and the inverse decay rate of unstable
particles.  The accurate estimations given in \cite{18_1} for a highly rotating Kerr black holes show that the lifetime of neutral pion $\tau (\pi_{0}) = 8 \times 10^{-17}$ s.
is of the same order as the e-folding time for the maximal instability growth  $\tau = 1.5 \times 10^{-17}$ s. Yet this happen only in the very narrow region of masses
of black holes $M \sim 10^{12} kg$. Assuming few times increasing of the instability growth rate in the presence of the magnetic field, we could expect much stronger
instability for neutral pion and much wider range of black hole masses for which such instability is essential.

\begin{figure}[htbp]
\centering
\includegraphics[scale=0.27]{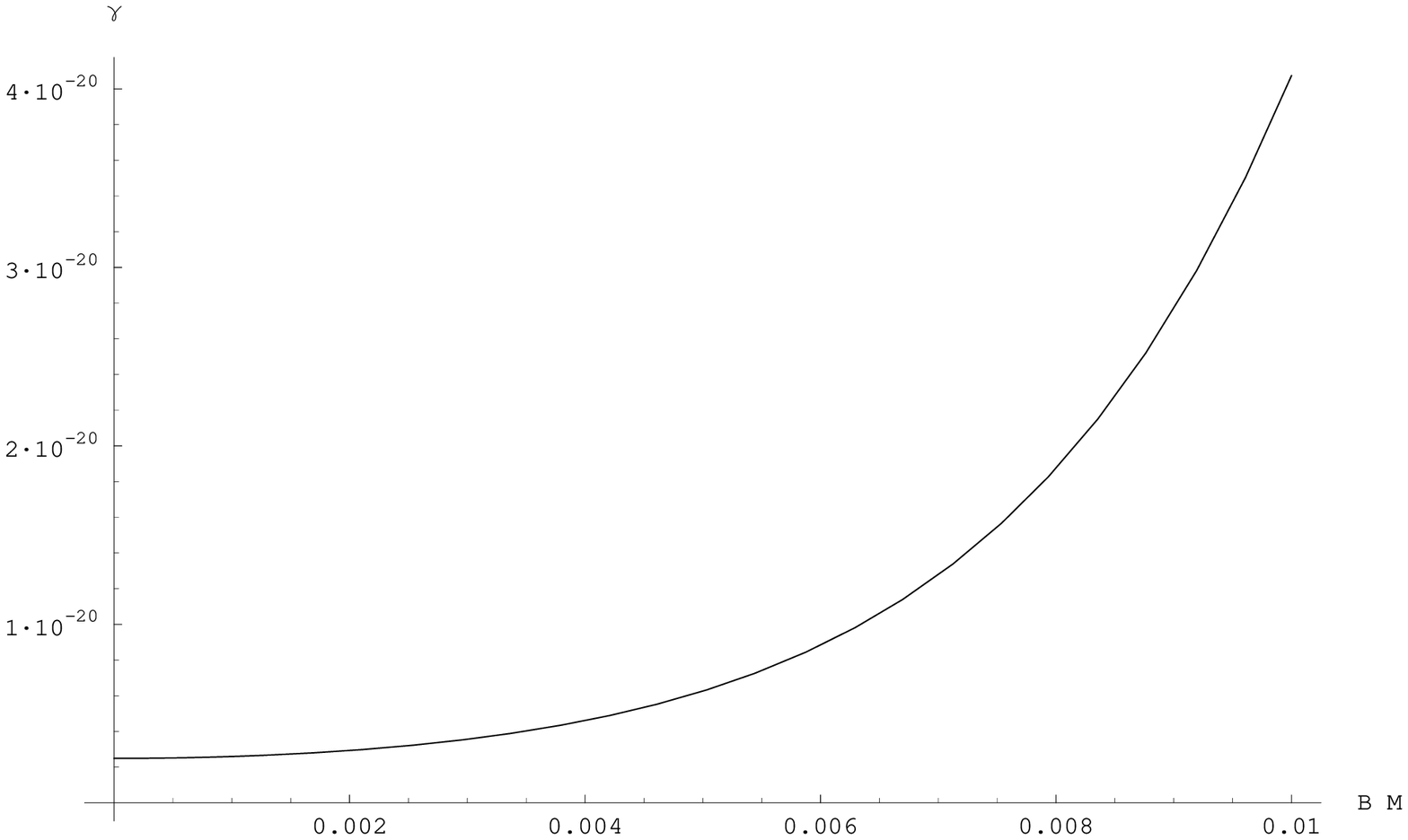},\includegraphics[scale=0.27]{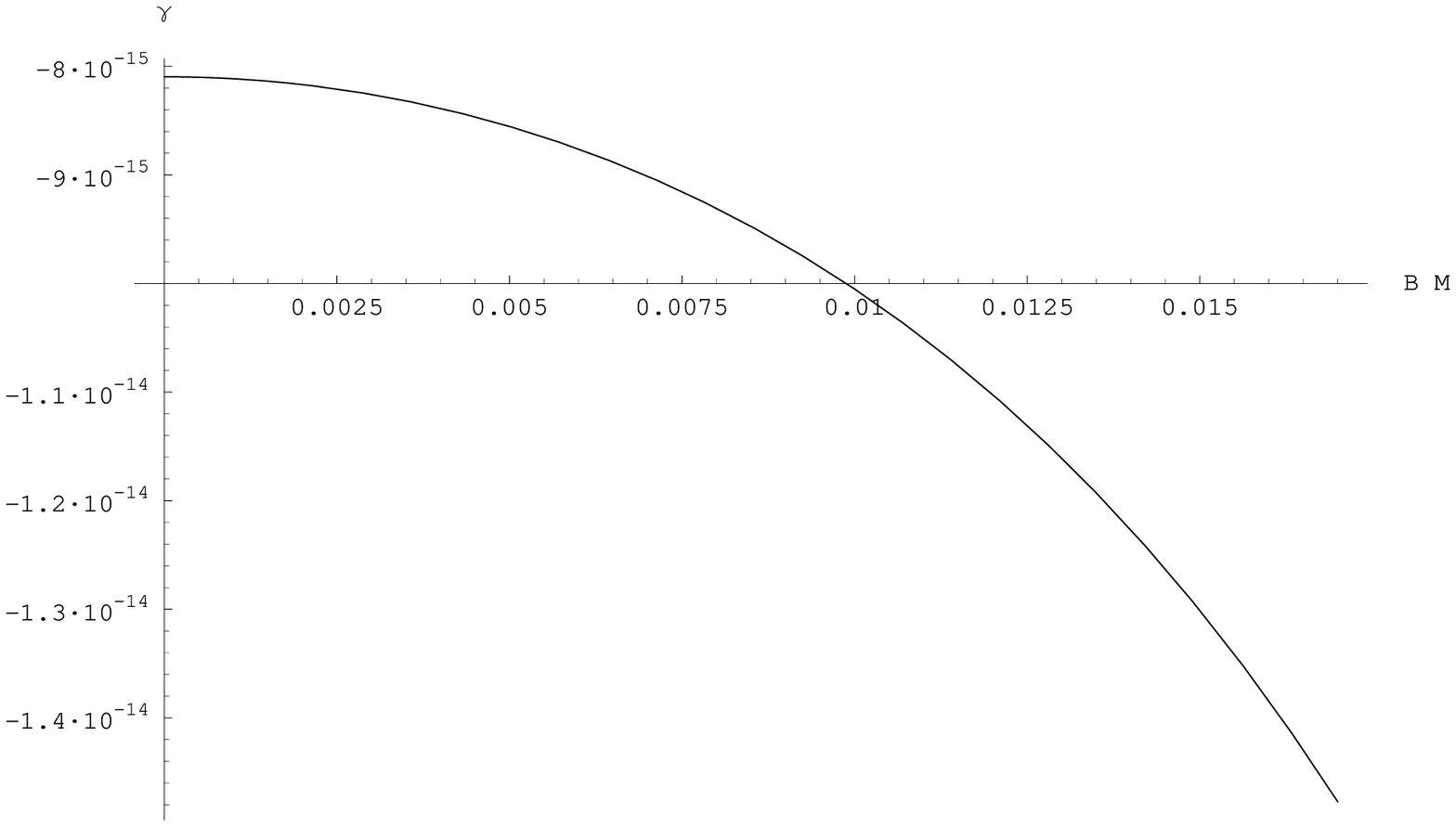},\includegraphics[scale=0.27]{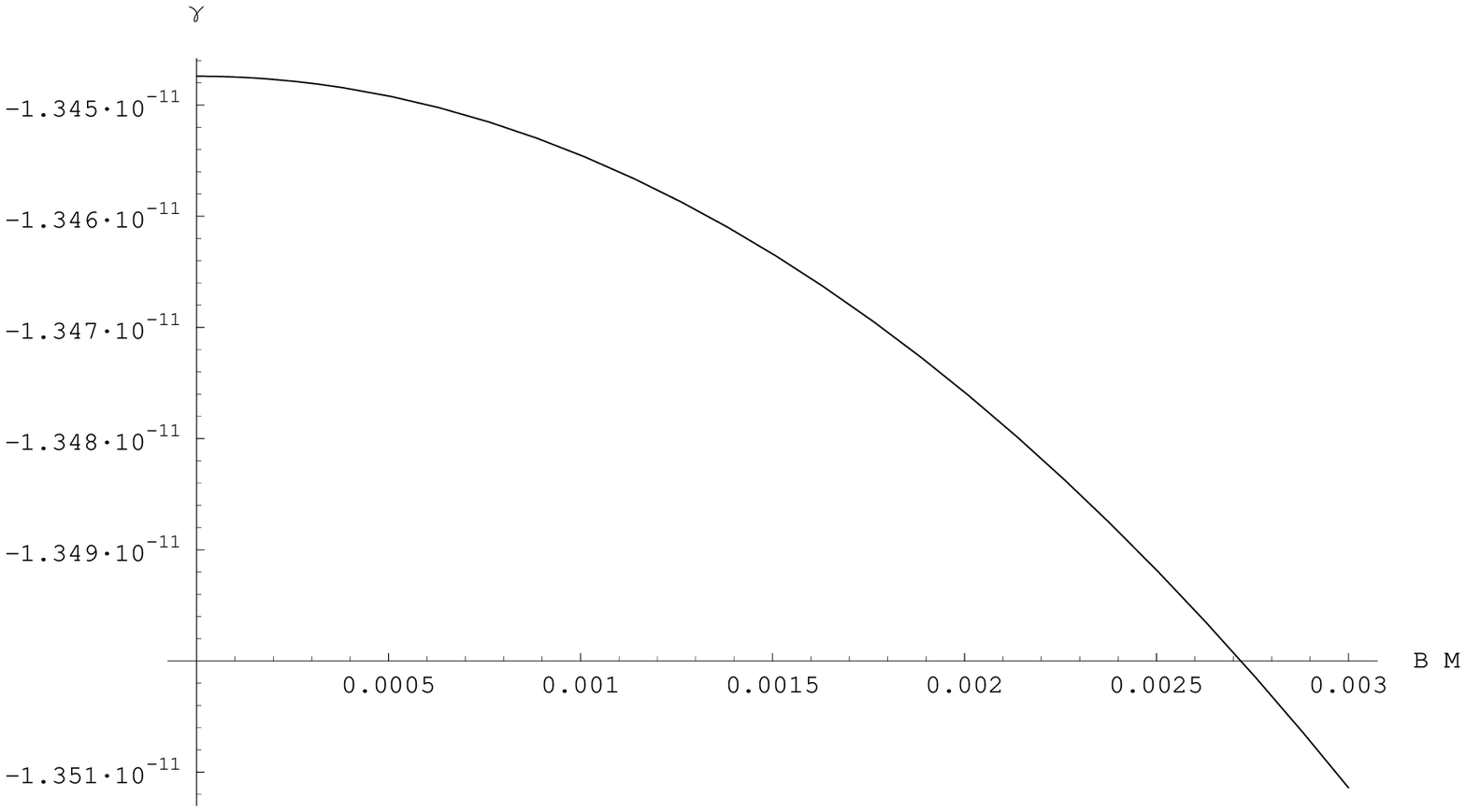}
\caption{The growth/damping rate $\gamma$ as a function of the dimensionless parameter $B M$ for massive scalar field: $\ell=m=1$, $n=0$, $a=0.1$,
$\mu M =0.1$ (left), $\mu M =0.05$, $\mu M =0.01$ (right). One can see that there is no superradiant instability for cases $\mu =0.1$ and $0.05$ as for those beyond the inequality (20). The superradiant instability growth rate, plotted in the right figure, is shown to be increased by a few times because of the presence of the magnetic field.}
\label{1}
\end{figure}

\section{Quasinormal modes}

Let us discuss here the propagation of the both massive and massless scalar fields with another boundary conditions, called
{\it the quasinormal} boundary conditions \cite{BC}. These are pure in-going waves at the event horizon and pure out-going at the spatial infinity.
\begin{equation}\nonumber
\Psi \sim e^{-i \omega r_{*}}, \quad r_{*} \rightarrow - \infty, \qquad and \qquad \Psi \sim e^{i \omega r_{*}} , \quad r_{*} \rightarrow + \infty,
\end{equation}
The absence of the waves incident on a potential barrier of the black hole gives rise to the formal absence of superradiance. Therefore one cannot
see  superradiant instability under pure quasinormal boundary conditions. After all, as it is clear from the previous section, the instability is
significant only for a narrow class of particles and black hole masses. The quasinormal modes of the considered here slowly rotating magnetized black holes
reduce to the quasinormal modes of Kerr black holes with "re-defined" mass of the field $\mu \rightarrow \sqrt{\mu^2 + B^2 m^2}$. It is essential that, similar to the scattering characteristics, the quasinormal modes also depend on the behavior of the effective potential near its peak, while behavior of the potential far from the black hole is insignificant \cite{KZ_PLB2}. This was proved in \cite{KZ_PLB2} by showing that quasinormal modes of two different potentials,
with equal behavior near the black hole and strikingly different behavior in the far region, coincide.  Using this fact, the quasinormal modes of non-rotating
magnetized black holes were found in \cite{14}. In a similar fashion, one can re-obtain the quasinormal modes of the slowly rotating black holes in
the magnetic field by using numerical data of \cite{Konoplya:2006br}-\cite{Ohashi:2004wr}  and by the re-definition $\mu \rightarrow \sqrt{\mu^2 + B^2 m^2}$.
Here we shall not repeat these stages  but refer a reader to the work \cite{Konoplya:2006br}. Here we simply enumerate the
main features of the quasinormal spectrum: First, $B |m|$  acting like a "mass", leads to decreasing of the imaginary part of the quasinormal
mode until.

The greater $B |m|$, the less the damping rate of the mode and for sufficiently large  value of $\sqrt{\mu^2 + B^2 m^2}$, the infinitely long lived
modes, called quasi-resonances \cite{Konoplya:2004wg}, may appear. Let us remind that we separated varibles under the condition $\mu M \ll 1$, so that
the existence of the quasi-resonances (which happen when $\mu M \geq \sim  1$) are guaranteed only for massless  scalar field in a
magnetized black hole background. Similar massless scalar field quasi-resonances  were found for non-rotating magnetized black holes in \cite{14} and
originally, for the Schwarzschild black holes in \cite{Ohashi:2004wr}. Second, in the regime of infinitely high damping, the quasinormal modes do not depend on magnetic field and are the same as for the massless scalar field in the pure Kerr background.
Finally, the real oscillation frequency $Re \omega$ decreases when increasing $B m$ until some minimal value and then increases until reaching
its quasi-resonance value.

Here we see that the perturbations under pure quasinormal boundary conditions are longer lived in the presence of the magnetic field. Unlike quasinormal modes,
scattering boundary conditions lead to the increasing of the damping rate of each mode in the stable sector (see Fig.2). We know that the
superradiant instability has the tiny growth rate, except some special cases of the neutral pion decay, which we discussed in the previous section.
All this does not mean any contradiction. First of all, because quasinormal modes are beyond the region of superradiant instability.
In addition, we implied here that, putting aside those exceptional cases, such an instability (and more general, superradiance) does not
influence the quasinormal modes. Although any realistic perturbation contains some remnants of the incident wave of the initial perturbative process, at
sufficiently late time the effect of this remnant is negligible.

\section{Conclusion}

We have separated variables for the Klein-Gordon and Hamilton-Jacoby equations in the limit $B \ll M^{-1}$. This corresponds to the "weak" magnetic field as for the deformation the black hole space-time, but a super-strong field in the astrophysical context. It is shown that the growth rate of the superradiant instability may be amplified by a few times due-to the magnetic field. This enlarges the range of black hole masses for which the instability may be significant in the neutral pion decay. The essential problem which was beyond our investigation
is the calculation of the instability growth for the highly rotating black holes. Another important question is a detailed description of the process of filling of the quasi-bound states \cite{Galtsov2}.

\section*{Appendix: Classical motion of particles near the magnetized black hole}

The equatorial motion of particles in the background of the Ernst space-time was
considered in \cite{Galtsov1}.
The Ernst matric can be obtained if one takes
$a=0$ limit in the Diaz metric (7). The Hamilton-Jacoby equation takes the form:
\begin{equation}
g_{\mu \nu} \frac{\partial S}{\partial x^{\mu}}\frac{\partial S}{\partial x^{\nu}} = \mu^2
\end{equation}
Implying that $S = -E t + L \phi + S_{1}(r, \theta)$, one has
\begin{equation}
\frac{r^2}{\Delta} E^2 - \frac{\Delta}{r^2} \left(\frac{\partial S_{1}}{\partial r}\right)^{2} -
\frac{1}{r^2} \left(\frac{\partial S_{1}}{\partial \theta}\right)^{2} -
\frac{\Lambda^4}{r^2 \sin^2 \theta} L^2 = \mu^2 \Lambda^2.
\end{equation}
Here we shall not be limited by equatorial plane but will imply that $B  \ll M^{-1}$ and $\mu^2 B^2 \ll 1$,
so that terms of order $\mu^2 B^2 \ll 1$ and higher will be discarded. Thus we replace $\Lambda^4$
on the left hand side of equation (22) by $1 + B^2 r^2 \sin^2 \theta$, and $\Lambda^2$ by unity on the right hand side.

Then the effective potential takes the same form as a that for the Schwarzschild black hole with shifter angulat momentum $L^2 \rightarrow L^2 (1 + B^2)$,
\begin{equation}
U_{eff}^2 =  \frac{\Delta \mu^2}{r^2} \left(1 + \frac{L^2 (1 + B^2)}{\mu^2 r^2} \right).
\end{equation}

Our main interest here is to determine parameters region when bound states, that is, classically stable orbits are possible.
One can easily see that minimal distance $r = 6 M$ at which the stable circular orbit exists is unchanged in the lowest
dominant order which we considered. Yet the minimal value of the angular momentum now depends on $B$
$$
\frac{L_{min} \sqrt{1 + B^2}}{M} = 3.464.
$$
In the equatorial plane the Hamilton-Jacoby equation (22) allows separation of variable without any approximation \cite{Galtsov1} and
one can obtain the exact $B$ - correction to the position of the stable circular orbit.

\section*{Acknowledgments}
This work was supported by the {\it Japan Society for the Promotion of Science (JSPS)}, Japan.
I would like to thank Taichi Koboyashi for useful discussions.

\end{document}